\documentclass[12pt, preprint]{aastex}
\newcommand{\Ibar}{I\hspace{-0.4em}\rule[0.35em]{0.3em}{0.02em}\hspace{0.1em}}
\shorttitle{Secular Bar-mode Instability in Neutron Stars}

\shortauthors{Ou, Tohline \& Lindblom}

\begin{document}


\title{{Nonlinear Development of the Secular Bar-mode Instability\\
in Rotating Neutron Stars}}


\author{Shangli Ou and Joel E. Tohline}
\affil{Department of Physics \& Astronomy, Louisiana State
University, Baton Rouge, LA  70803}

\and

\author{Lee Lindblom}
\affil{Theoretical Astrophysics 130-33,
       California Institute of Technology, Pasadena, CA 91125}


\begin{abstract}
We have modelled the nonlinear development of the secular bar-mode
instability that is driven by gravitational radiation-reaction
(GRR) forces in rotating neutron stars.  In the absence of any
competing viscous effects, an initially uniformly rotating,
axisymmetric $n=1/2$ polytropic star with a ratio of rotational to
gravitational potential energy $T/|W| = 0.181$ is driven by GRR
forces to a bar-like structure, as predicted by linear theory. The
pattern frequency of the bar slows to nearly zero, that is, the
bar becomes almost stationary as viewed from an inertial frame of
reference as GRR removes energy and angular momentum from the
star.  In this ``Dedekind-like'' state, rotational energy is
stored as motion of the fluid in highly noncircular orbits inside
the bar. However, in less than 10 dynamical times after its
formation, the bar loses its initially coherent structure as the
ordered flow inside the bar is disrupted by what appears to be a
purely hydrodynamical, short-wavelength, ``shearing'' type
instability.  The gravitational waveforms generated by such an
event are determined, and an estimate of the detectability of
these waves is presented.
\end{abstract}


\begin{deluxetable}{lllcll}
\tablecolumns{6}
\tablewidth{0pt}
\tablecaption{Initial Model Parameters\label{TabInitial}}
\tablehead{
\colhead{} & \multicolumn{2}{c}{Model SPH} &
&\multicolumn{2}{c}{Model ROT181} \\
\cline{2-3}\cline{5-6}
\colhead{Units} & \colhead{code} & \colhead{cgs} && \colhead{code} &
\colhead{cgs}
}
\startdata
$c$ & 3.122 & $3.00\times 10^{10}$ && 1.634 & $3.00\times 10^{10}$\\
$M$ & 1.359 & $2.80\times 10^{33}$ && 0.1716 & $2.80\times 10^{33}$\\
$K$ & 0.7825 & $1.83\times 10^{-10}$ && 0.1281 & $1.83\times 10^{-10}$ \\
$R_{\mathrm{eq}}$
& 0.8413 & $1.25\times 10^6$ && 0.6102 & $1.97\times 10^{6}$\\
$R_{\mathrm{pole}}$
& 0.8413 & $1.25\times 10^6$ && 0.2746 & $8.86\times 10^{5}$\\
$\bar\rho$ & 0.5460 & $3.42\times 10^{14}$ && 0.4922 & $2.39\times 10^{14}$\\
$\Omega_{\mathrm{rot}}$ & 0.0 & 0.0 &&0.9705 & $5.52\times 10^3$\\
$J$ & 0.0 & 0.0 && 0.01632 & $1.58\times 10^{49}$
\enddata
\end{deluxetable}

\section{Introduction}

As \cite{Ch69} and \cite{T78} have discussed in depth, if a star
rotates sufficiently fast -- to a point where the ratio of
rotational to gravitational potential energy in the star $T/|W|
\gtrsim 0.27$ -- it will encounter a dynamical instability that
will result in the deformation of the star into a rapidly
spinning, bar-like structure.  Although originally identified in
configurations that are uniformly rotating and uniform in density,
more generalized analyses made it clear that the dynamical
bar-mode instability should arise at approximately the same
critical value of $T/|W|$ in centrally condensed and
differentially rotating stars (cf., Ostriker \& Bodenheimer 1973).
A number of groups have used numerical hydrodynamic techniques to
follow the nonlinear development of this bar-like structure in the
context of the evolution of protostellar gas clouds \citep{TDM85,
DGTB86, WT88, PCDL98,CT00} and in the context of rapidly rotating
neutron stars \citep{NCT00,B00}. Very recently numerical
simulations by \cite{CNLB01} and \cite{SKE02,SKE03} have shown
that low-order nonaxisymmetric instabilities can become
dynamically unstable at much lower values of $T/|W|$ in stars that
have rather extreme distributions of angular momentum. Through
linear stability analyses, \cite{KE03} and \cite{WAJ04} have
attempted to show the connection between these instabilities and
the classical bar-mode instability discovered in stars with less
severe distributions of angular momentum.  In our present analysis
we will not be directly investigating the onset or development of
these dynamical instabilities.

Classical stability studies have also indicated that a uniformly
rotating (or moderately differentially rotating) star with $T/|W|
\gtrsim 0.14$ should encounter a secular instability that will
tend to deform its structure into a bar-like shape if the star is
subjected to a dissipative process capable of redistributing
angular momentum within its structure. The nonlinear development
of this secular instability has not previously been modelled in a
fully self-consistent manner, so it is not yet clear whether stars
that encounter this type of instability will evolve to a structure
that has a significant bar-like distortion. In this paper, we
present results from a numerical simulation that has been designed
to follow the nonlinear development of the secular bar-mode
instability in a rapidly rotating neutron star.  A force due to
gravitational radiation-reaction (GRR) serves as the dissipative
mechanism that drives the secular development of the bar-mode.  By
following the development of the bar to a nonlinear amplitude and
calculating the rate at which angular momentum and energy are lost
from the system due to gravitational radiation, we are able to
provide a quantitative estimate of the distance to which such a
gravitational-wave source could be detected by existing and
planned experiments, such as the laser interferometer
gravitational-wave observatory (LIGO).

\cite{Ch70} was the first to discover that gravitational
radiation-reaction forces can excite the secular bar-mode
instability in uniformly rotating, uniform-density stars with
incompressible equations of state ({\it i.e.}, the Maclaurin
spheroids). This work was generalized by \cite{FS78} and
\cite{C79a,C79b}, to show that the GRR instability extends to
stars with any equation of state, and to other nonaxisymmetric
modes with higher azimuthal mode numbers ($m>2$). \cite{M85},
\cite{IFD85}, \cite{IL90}, and \cite{LS95} have all shown that the
critical value of $T/|W|$ at which the GRR secular instability in
the ($m=2$) bar-mode sets in does not depend sensitively on the
polytropic index of the equation of state or the differential
rotation law of the star. These stability analyses have also been
generalized to systems in which the star is governed by
relativistic, rather than purely Newtonian, hydrodynamics and
gravitational fields \citep{F78,LH83,C91,CL92,SF98,SZ98,DiGV02}.

\cite{LD77} first showed that viscous processes within compact stars
can act to suppress the GRR-driven secular bar-mode instability.
Various physical viscosities have been considered, for example, shear
viscosity due to electron and neutron scattering \citep{FI76,CL87},
shear viscosity due to neutrino scattering \citep{KS77,LD79,TD93},
bulk viscosity due to weak nuclear interactions
\citep{J71,S89,IL91,YE95}, or ``mutual friction'' effects in a
superfluid \citep{LM95}. In the present work, we will not be
investigating the influence of viscous processes on the GRR-driven
bar-mode instability, focusing instead on following the nonlinear
development of the bar-mode in purely inviscid systems.

\cite{DL77} and \cite{LS95} have made efforts to follow the
non-linear evolution of rotating neutron stars that are
susceptible to the secular (or the dynamical) bar-mode instability
by using energy and angular momentum conservation to construct a
sequence of quasi-equilibrium, ellipsoidal configurations.  Here,
we follow the GRR-driven evolution of the bar-mode in an even more
realistic way by integrating forward in time the coupled set of
nonlinear partial differential equations that govern dynamical
motions in nonrelativistic fluids, and by including a
post-Newtonian radiation-reaction force term in the equation of
motion. Reviews of this and other instabilities that are expected
to arise in young neutron stars have been written by \cite{L97},
\cite{L01}, \cite{S03}, and \cite{A03}.

\section{Methods}
Using the \cite{H86} self-consistent-field technique, we
constructed two initial equilibrium stellar models governed by
Newtonian gravity and an $n=1/2$ polytropic equation of state;
that is, the gas pressure $p$ and density $\rho$ were related
through the expression $p = K\rho^{1+1/n}$, where $K$ is a
constant. Model ``SPH'' was initially nonrotating and, hence,
spherically symmetric; model ``ROT181'' was initially
axisymmetric, and uniformly rotating with a ratio of rotational to
gravitational potential energy $T/|W| = 0.181$. Other properties
of these two initial models are detailed in Table
\ref{TabInitial}: $M$ is the mass of the star; $R_{\mathrm{eq}}$
and $R_{\mathrm{pole}}$ are the star's equatorial and polar radii,
respectively; $\bar{\rho}$ is the star's mean density;
$\Omega_{\mathrm{rot}}$ is the angular velocity of rotation; and
$J$ is the star's total angular momentum. Columns 2 and 4 of Table
\ref{TabInitial} give the values of these various quantities in
dimensionless code units, where we have assumed the gravitational
constant, the star's central density, and the radial extent of the
computational grid are all equal to unity ({\it i.e.},
$G=\rho_{\mathrm{c}} = \varpi_{\mathrm{grid}}=1$). Columns 3 and 5
of Table \ref{TabInitial} give the value of each quantity in cgs
units, assuming both stars have $M =1.4~\mathrm{M}_{\odot}$ and $K
= 1.83\times10^{-10} ~\mathrm{cm^8~g^{-2}~s^{-2}}$. (This value of
$K$ produces a spherical $1.4~\mathrm{M}_{\odot}$ star with a
radius of $12.5~\mathrm{km}$, which is characteristic of a neutron
star.)

\begin{figure}
\epsscale{1.0}\plotone {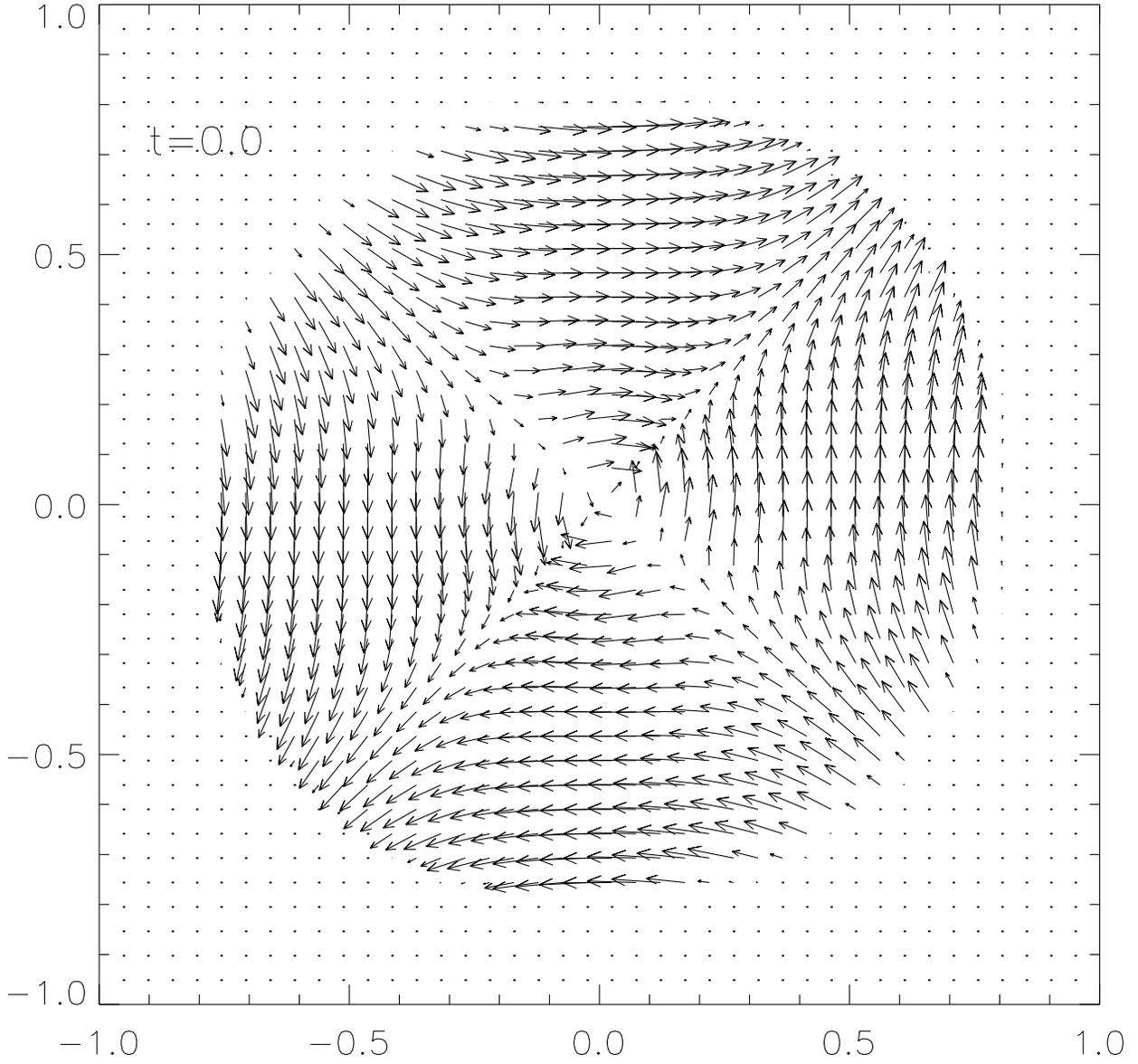} \caption{Velocity vectors in the
equatorial plane of model SPH at time $t=0$, but after the
nonrotating model was perturbed by the $\ell=m=2$ ``bar-mode''
eigenfunction drawn from linear theory.\label{velPerturb}}
\end{figure}

Each model was introduced into our hydrodynamical code along with
a low-amplitude, nonaxisymmetric perturbation that was designed to
closely approximate the eigenfunction of the $\ell=m=2$
``bar-mode'' in a spherical, $n=1/2$ polytrope \citep{IL90}. As an
illustration, Fig.~\ref{velPerturb} shows the perturbed velocity
field that was introduced in the equatorial plane of model SPH
along with a low-amplitude, bar-like distortion in the density
which oriented the bar along the vertical axis. Then the nonlinear
hydrodynamical evolution of each model was followed using the
numerical simulation techniques described in detail by
\cite{MTF02}. More specifically, we integrated forward in time a
finite-difference approximation of the following coupled set of
partial differential equations:
\begin{eqnarray}
\frac{\partial\rho}{\partial t}+\vec{\nabla}\cdot(\rho\vec{v})&=&0,\\
\label{motion}\rho(\frac{\partial\vec{v}}{\partial t} +
\vec{v}\cdot\vec{\nabla}\vec{v}) &=& -\vec{\nabla}p
- \rho\vec{\nabla}(\Phi + \kappa \Phi_{GR}),\qquad\\
\frac{\partial{\tau}}{\partial{t}}+\nabla\cdot(\tau \vec{v})&=&0,\\
\nabla^2\Phi &=& 4\pi G\rho,
\end{eqnarray}
where $\vec{v}$ is the fluid velocity, \($$\Phi$$\) is the
Newtonian gravitational potential, $\tau \equiv
(\epsilon\rho)^{1/\Gamma}$ is the entropy tracer, $\epsilon$ is
the specific internal energy, $p=$$(\Gamma-1)\epsilon\rho$ and
$\Gamma=1+1/n =3$. Because the models were initially constructed
using a polytropic index $n=1/2$ and they were evolved using an
adiabatic form of the first law of thermodynamics (Eq. 3) with an
adiabatic exponent $\Gamma=3$, the models effectively maintained
uniform specific entropy at a value specified by the initial
model's polytropic constant, $K$.

In the equation of motion, Eq.~(2), we included
the post-Newtonian approximation to the gravitational
radiation-reaction potential produced by a time-varying,
$\ell=m=2$ mass-quadrupole moment \citep{IL91},
\begin{equation}
    \Phi_{GR} \equiv -\sqrt{\frac{2\pi}{375}} \biggl(\frac{G}{c^5}\biggr)\varpi^2e^{2i\phi}D_{22}^{(5)}  \,
    ,
\end{equation}
where $D_{22}^{(5)}$ is the fifth time-derivative of the
quadrupole moment and $c$ is the speed of light. For modeling
purposes, a dimensionless coefficient $\kappa$ was affixed to the
radiation-reaction potential term in the equation of motion. By
adjusting the value of $\kappa$, we could readily remove or
artificially enhance the effect of this non-Newtonian GRR force.

As implemented on our cylindrical computational mesh
($\varpi,\phi,z$), $D_{22}$ and its first time-derivative were
evaluated using the following expressions,
\begin{eqnarray}
    D_{22}&=&\sqrt{\frac{15}{32\pi}}\int\rho\varpi^2e^{-2i\phi}d^{\,3}x \, , \\
    D_{22}^{(1)}&=&\sqrt{\frac{15}{8\pi}}\int\rho\varpi[v_{\varpi}-iv_{\phi}]e^{-2i\phi} d^{\,3}x \, .
\end{eqnarray}
Following Lindblom, Tohline, \& Vallisneri (2001, 2002) we have
assumed that the quadrupole moment has a time-dependence of the
form $D_{22}\propto e^{-i\omega_{22}t}$, hence,
\begin{equation}
D_{22}^{(n)} = (-i\omega_{22})^n D_{22} \, ,
\end{equation}
where, at any instant in time, the complex eigenmode frequency
$\omega_{22}=\omega_{\mathrm{r}} + i\omega_{\mathrm{i}}$ can be
determined by taking the ratio $D_{22}^{(1)}/D_{22}$, so in
Eq.~(5) we set
\begin{equation}
D_{22}^{(5)} = |\omega_{22}|^4 D_{22}^{(1)} \, \label{eqn:D22_d5} .
\end{equation}

\section{Predictions of Linear Theory}

\begin{figure}
\epsscale{.90}\plotone{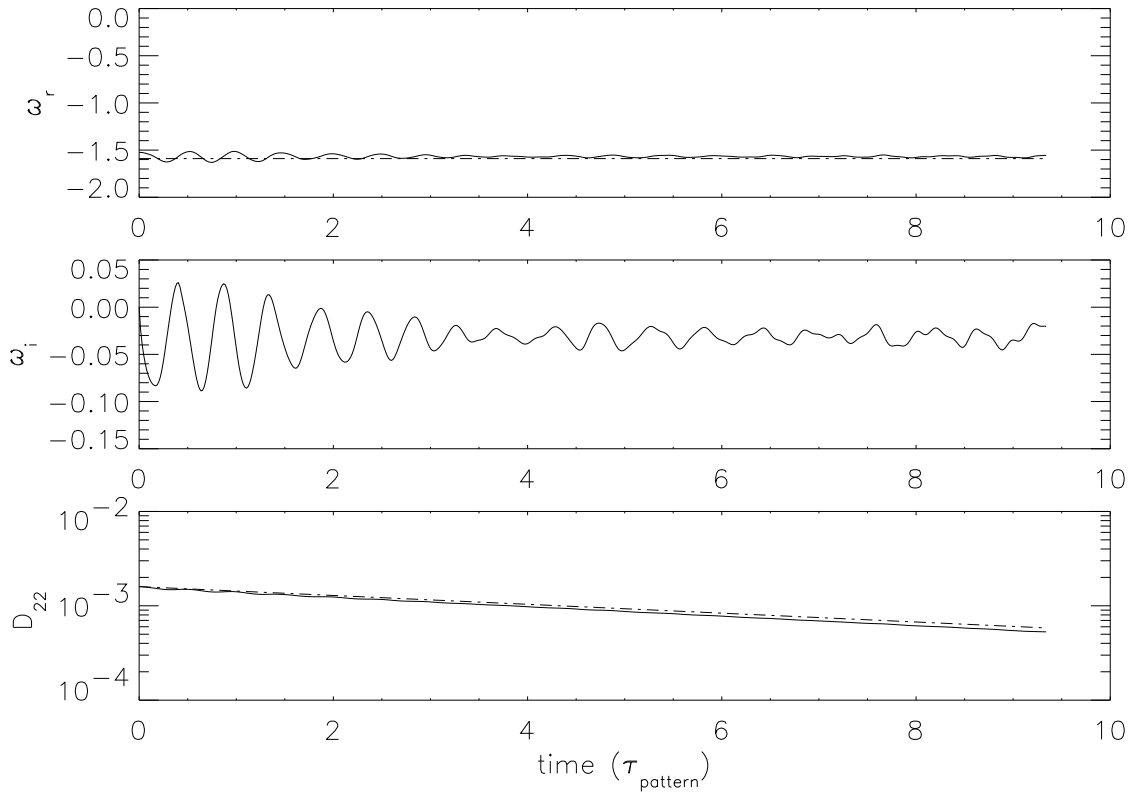} \caption{Solid curves depict the
time-evolution of the amplitude $|D_{22}|$ (bottom) and the real
(top) and imaginary (middle) components of the $\ell=m=2$ bar-mode
frequency from model SPH. Time is shown in units of the predicted
pattern period; $|D_{22}|$ has been normalized to
$MR_{\mathrm{eq}}^2$; and frequencies are shown in dimensionless
code units. Dash-dotted lines show predictions of linear theory.
\label{SPHfrequencies}}
\end{figure}

Using the linear perturbation techniques described by
\cite{IL90,IL91} we determined that in model SPH,
$\omega_{\mathrm{r}} = \mathrm{Re}(\omega_{22}) = \pm (1.21 \pm
0.01)\Omega_0$ and, if the model is scaled to a mass of $1.4
M_{\odot}$ and a radius of 12.5 km, $\omega_{\mathrm{i}}=
\mathrm{Im}(\omega_{22}) = -
1.00\pm0.01\times10^{-3}\kappa\Omega_0$, where $\Omega_0 \equiv
\sqrt{\pi G\bar{\rho}} = 1.308$ in dimensionless code units. (The
uncertainty is estimated from the values that are determined
numerically by the linear perturbation method for different radial
resolutions.) In model SPH, therefore, we should expect the
amplitude of the bar-mode to damp exponentially on a time scale
$\tau_{\mathrm{GR}} \equiv 1/|\omega_{\mathrm{i}}| = 193
\kappa^{-1}\tau_{\mathrm{pattern}}$, where
$\tau_{\mathrm{pattern}}\equiv 2\pi/|\omega_{\mathrm{r}}| = 3.97$
in dimensionless code units.

Linear perturbation analyses have not yet provided quantitative
values of the bar-mode eigenfrequency in rapidly rotating, $n=1/2$
polytropes.  From the information given in \cite{IL90,IL91},
however, we expect that in model ROT181, (a)
$\omega_{\mathrm{r}}/\Omega_0$ should be positive but close to
zero, as viewed from an inertial reference frame; and (b)
$\omega_{\mathrm{i}}/\Omega_0$ should be slightly positive -- that
is, the mass-quadrupole moment should {\it grow} exponentially,
but on a time scale that is very long compared to the {\it
damping} time predicted for the nonrotating model, SPH. More
specifically, if $|\omega_{\mathrm{r}}|/\Omega_0$ proves to be an
order of magnitude smaller in model ROT181 than it is in model
SPH, then we should expect $\tau_{\mathrm{GR}}$ to be $\sim 10^5$
times larger because the amplitude of the GRR driving term in Eq.
(2) is proportional to $\omega_{22}^5$.

\section{Barmode Evolutions}
\subsection{Model SPH}

Initially the perturbation applied to model SPH had an amplitude
$D_0 \equiv |D_{22}(t=0)| \approx 10^{-3}$ and a velocity field
(see Fig.~\ref{velPerturb}) designed to excite the ``backward
moving'' $\ell=m=2$ bar-mode, that is, the mode for which
$\omega_{\mathrm{r}} < 0$. We followed the evolution of the model
on a cylindrical grid with a resolution of $66 \times 128 \times
130$ zones in $\varpi$, $\phi$, and $z$, respectively, and with
the coefficient of the radiation-reaction force term in Eq.~(2)
set to the value $\kappa = 20$. The effect of this was to shorten
the timescale for the exponential decay by a factor of 20, to a
predicted value $\tau_{\mathrm{GR}} = 9.63
\tau_{\mathrm{pattern}}$. By shortening the decay timescale in
this manner, we were able to significantly reduce the amount of
computational resources that were required to follow the decay of
the barmode while maintaining a decay rate that was slow compared
to the characteristic dynamical time of the system, $\Omega_0^{-1}
\approx 0.2\tau_{\mathrm{pattern}}$. This is the same technique
that was successfully employed by \cite{LTV01,LTV02} in an earlier
investigation of the r-mode instability in young neutron stars.

Figure \ref{SPHfrequencies} displays the key results from our SPH
model evolution.  The solid curves in the top two frames display
$\omega_{\mathrm{r}}$ and $\omega_{\mathrm{i}}$ as a function of
time through just over nine pattern periods.  Each of these
frequencies oscillate about a fairly well-defined, mean value:
$\langle\omega_{\mathrm{r}}\rangle\approx - 1.56 = -
1.19\Omega_0$; $\langle\omega_{\mathrm{i}}\rangle\approx -0.03 =
-0.023\Omega_0$. Oscillations about these mean values initially
had an amplitude $\sim \pm 0.05 = 0.038\Omega_0$, indicating that
our initial nonaxisymmetric perturbation did not excite a pure
eigenmode, but these oscillations decreased in amplitude somewhat
as the evolution proceeded.  Our measured values of
$\omega_{\mathrm{r}}$ and $\omega_{\mathrm{i}}$ are within 3\% and
15\%, respectively, of the values predicted from linear theory
(see \S3).  The solid curve in the bottom frame of
Fig.~\ref{SPHfrequencies} shows in a semi-log plot the behavior of
$|D_{22}|$ with time.  (Note that in this figure, $|D_{22}|$ has
been normalized to $MR_{\mathrm{eq}}^2$.)  There is a clear
exponential decay with a measured damping time (given by the slope
of the solid curve) $\tau_{\mathrm{GR}}\approx
8.45\tau_{\mathrm{pattern}}$. This decay time is completely
consistent with the measured value of
$\langle\omega_{\mathrm{i}}\rangle$ that we have obtained from the
middle frame of Fig.~\ref{SPHfrequencies} and, again, within 15\%
of the predicted value (illustrated by the solid dash-dotted line
in the top frame of the figure).  The somewhat larger discrepancy
in the measured value of $\omega_i$ is most probably due to the
fact that the GRR formalism used here was derived under the
assumption that $|\omega_i|<<|\omega_r|$.  Since $\omega_i$ is
caused by the $\Phi_{GR}$ potential which is proportional to the
fifth power of the frequency, fractional discrepancies which are
of order $5|\omega_i/\omega_r|\approx 0.1$ are not unexpected.

The first row of numbers in Table \ref{TabResults} summarizes
these simulation results.  Specifically, columns 4, 5 and 6 list
the values of $\omega_r$, $\omega_i$, and $\tau_{\mathrm{GR}}$
that have been drawn directly from Fig.~\ref{SPHfrequencies}; all
three numbers are given in dimensionless code units. In the last
two columns of this table, the real and imaginary frequencies have
been reexpressed in units of the dynamical frequency, $\Omega_0$.
In the last column, we also have adjusted $\omega_i$ by the factor
of $\kappa$ in order to show the frequency (and associated growth
rate) as it would appear in a real neutron star where the GRR
force would not be artificially exaggerated.

\subsection{Model ROT181}

\begin{deluxetable}{lrrrrcrc}
\tablewidth{0pt} \tablecaption{Simulation
Results\label{TabResults}}

\tablehead{ \colhead{Model} & \colhead{$\Omega_0$} &
\colhead{$\kappa$} & \colhead{$\omega_{\mathrm{r}}$} &
\colhead{$\omega_{\mathrm{i}}$} & \colhead{$\tau_{\mathrm{GR}}$} &
\colhead{$\omega_{\mathrm{r}}/\Omega_0$}  &
\colhead{$\omega_{\mathrm{i}}/(\Omega_0\kappa)$} } \startdata

SPH    & 1.308 & $2.00\times10^1$ &  $-1.56$ & $-0.03$ & 34 & $-1.19$ & $-1.1\times10^{-3}$ \\
ROT181 & 1.488 & $1.75\times10^5$ &  $+0.27$ & $+0.08$ & 12 & $+0.18$ & $+3.1\times10^{-7}$ \\

\enddata
\end{deluxetable}

\subsubsection{Radiation-reaction with $\kappa = 1.75\times10^5$}
Model ROT181 was introduced into our hydrocode with a
nonaxisymmetric perturbation in the density that had the same
structure as the perturbation that was introduced into model SPH.
Because we expected the natural oscillation frequency of the
bar-mode to be close to zero (as viewed from an inertial reference
frame), however, we did not perturb the velocity field of the
model. We followed the evolution of model ROT181 on a cylindrical
grid with a resolution of $130\times128\times98$ zones in
$\varpi$, $\phi$, and $z$, respectively, and with the coefficient
of the radiation-reaction force term set to $\kappa =
1.75\times10^5$.  Note that fewer vertical grid zones were
required than in model SPH because model ROT181 was significantly
rotationally flattened, but more radial zones were used than in
model SPH in order to allow room for model ROT181 to expand
radially during the nonlinear-amplitude phase of its evolution. A
much larger value of $\kappa$ was selected because, as explained
earlier, the natural growth rate of the bar-mode in model ROT181
was expected to be orders of magnitude smaller than the decay rate
measured in model SPH.

\begin{figure}
\epsscale{.90} \plotone{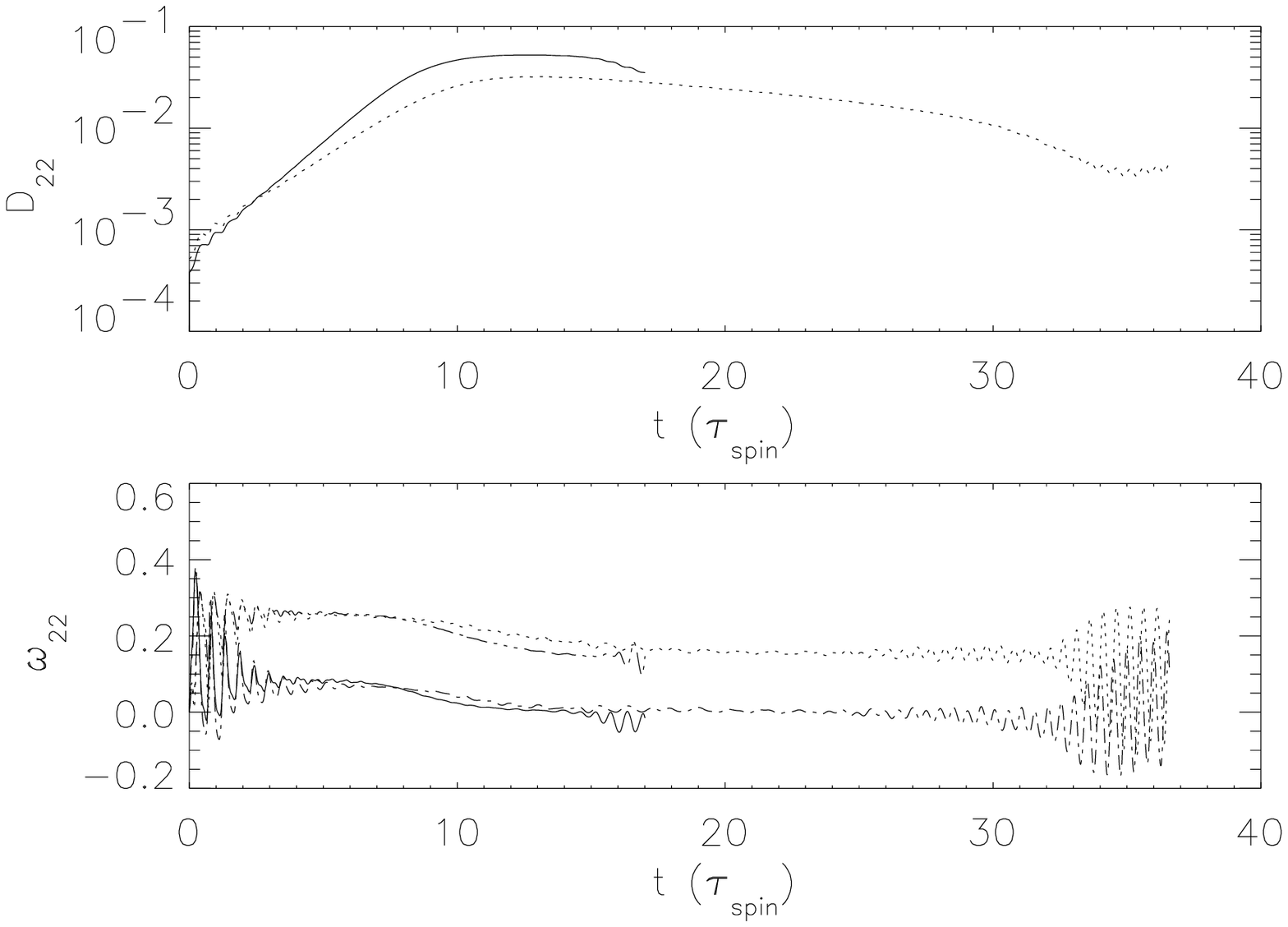} \caption{The time-evolution of the
amplitude $|D_{22}|$ (top) and the $\ell=m=2$ bar-mode frequency
$\omega_{22}$ (bottom) from two rapidly rotating models. Time is
shown in units of the initial rotation period
$\tau_{\mathrm{spin}}=2\pi/\Omega_{\mathrm{rot}}$ of the model;
$|D_{22}|$ has been normalized to $MR_{\mathrm{eq}}^2$; and
frequencies are shown in dimensionless code units. Curves that
terminate at approximately $17\tau_{\mathrm{spin}}$ display data
from model ROT181 and curves that extend past
$35\tau_{\mathrm{spin}}$ show data from the lower resolution model
ROT179. In the bottom panel, both the real (eg., dash-dotted curve
for model ROT181) and imaginary (eg., solid curve for model
ROT181) components of $\omega_{22}$ are displayed.
\label{rot181freq}}
\end{figure}

\begin{figure}
\epsscale{.90} \plotone{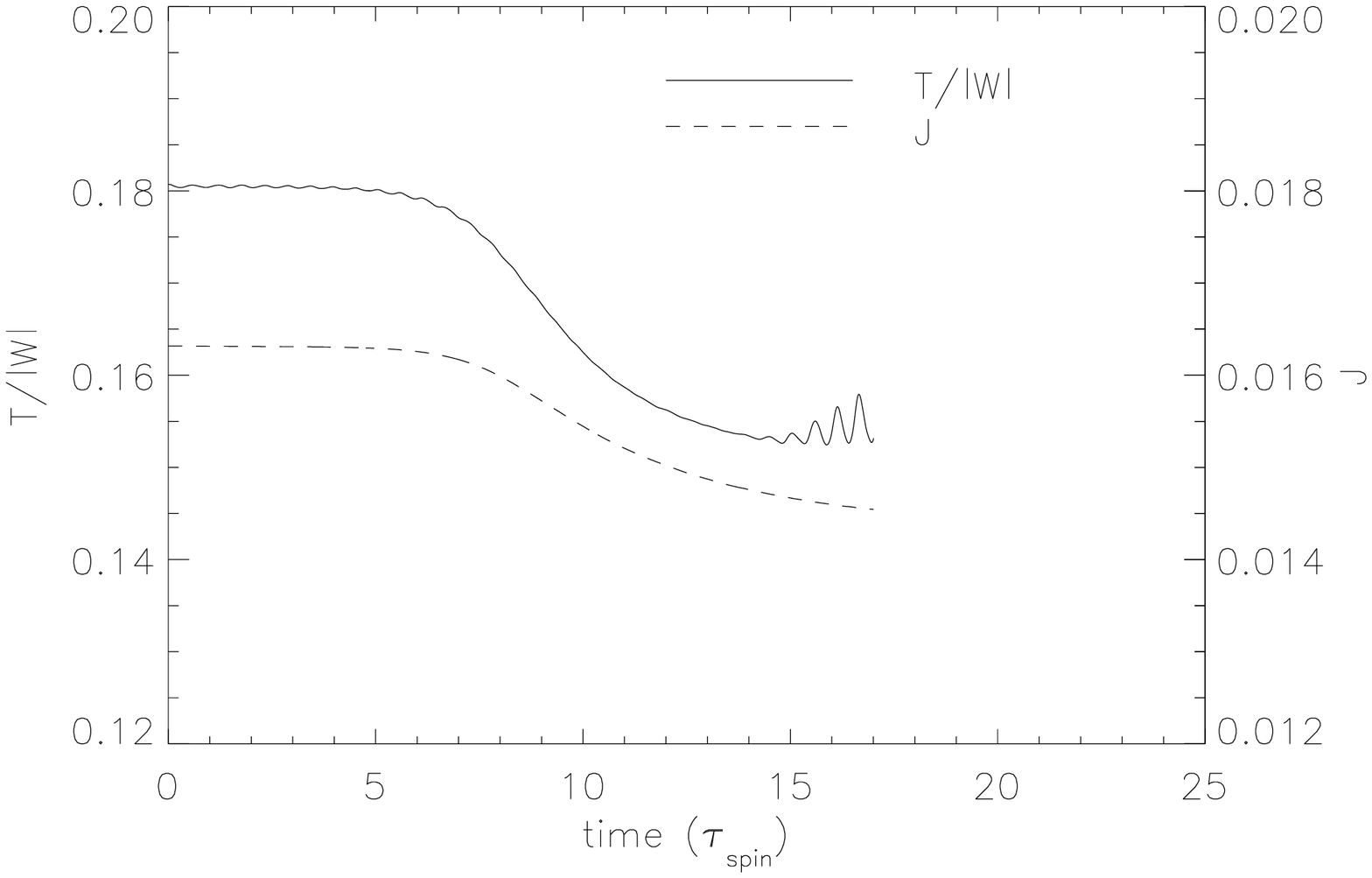} \caption{The time-evolution of the
angular momentum $J$ and the energy ratio $T/|W|$ from model
ROT181; $J$ is in dimensionless code units, time is shown in units
of the initial rotation period of the model. During the
{\it intermediate} phase of the evolution, both quantities noticeably
drop as angular momentum is lost via the GRR force term in the
equation of motion. \label{rot181ToverW}}
\end{figure}

Figures \ref{rot181freq} and \ref{rot181ToverW} display some of
the key results from this ROT181 model evolution.  The bottom
frame of Fig. \ref{rot181freq} shows the time-dependent behavior
of the real (dash-dotted curve) and imaginary (solid curve)
components of $\omega_{22}$, in our code's dimensionless frequency
units; the solid curve in the top frame displays the time-dependent behavior of
$|D_{22}|$, normalized to $MR_{\mathrm{eq}}^2$.  Figure
\ref{rot181ToverW} shows how the global parameters $T/|W|$ (solid
curve) and $J$ (dashed curve) evolve with time.  The behavior of
the model can be best described in the context of three different
evolutionary phases:  {\it Early} [$0 \leq t/\tau_{\mathrm{spin}}
\lesssim 7$]; {\it intermediate} [$7 \lesssim
t/\tau_{\mathrm{spin}} \lesssim 12$]; and {\it late}
[$t/\tau_{\mathrm{spin}} \gtrsim 12$], where
$\tau_{\mathrm{spin}}\equiv 2\pi/\Omega_{\mathrm{rot}} = 6.47$ in
dimensionless code units.

During the model's {\it early} evolution, both components of the
frequency $\omega_{22}$ oscillate about well-defined, mean values:
$\langle\omega_{\mathrm{r}}\rangle\approx 0.27 = 0.181\Omega_0$;
$\langle\omega_{\mathrm{i}}\rangle\approx 0.08 = 0.054\Omega_0$.
(Following Ipser \& Lindblom 1991, we define $\Omega_0$ in terms
of the mean density $\bar{\rho}_0$ of a spherical star that has
the same $M$ and $K$ as model ROT181, that is, $\Omega_0 \equiv
\sqrt{\pi G \bar{\rho}_0} = 1.488$ in dimensionless code units;
see Table \ref{TabResults}.) During this same phase of the
evolution, both $J$ and $T/|W|$ remain fairly constant, but
$|D_{22}|$ increases exponentially with a growth time (obtained
from the slope of the displayed curve) $\tau_{\mathrm{GR}}\approx
1.85 \tau_{\mathrm{spin}}$. This growth time is completely
consistent with the measured value of
$\langle\omega_{\mathrm{i}}\rangle$, from which we would expect
$\tau_{\mathrm{GR}}/\tau_{\mathrm{spin}} =
\langle\omega_{\mathrm{i}}\rangle^{-1}
(\Omega_{\mathrm{rot}}/2\pi) = 1.93$.  The second row of numbers
in Table \ref{TabResults} summarizes these simulation results.

\begin{figure}
\epsscale{1.0} \plotone{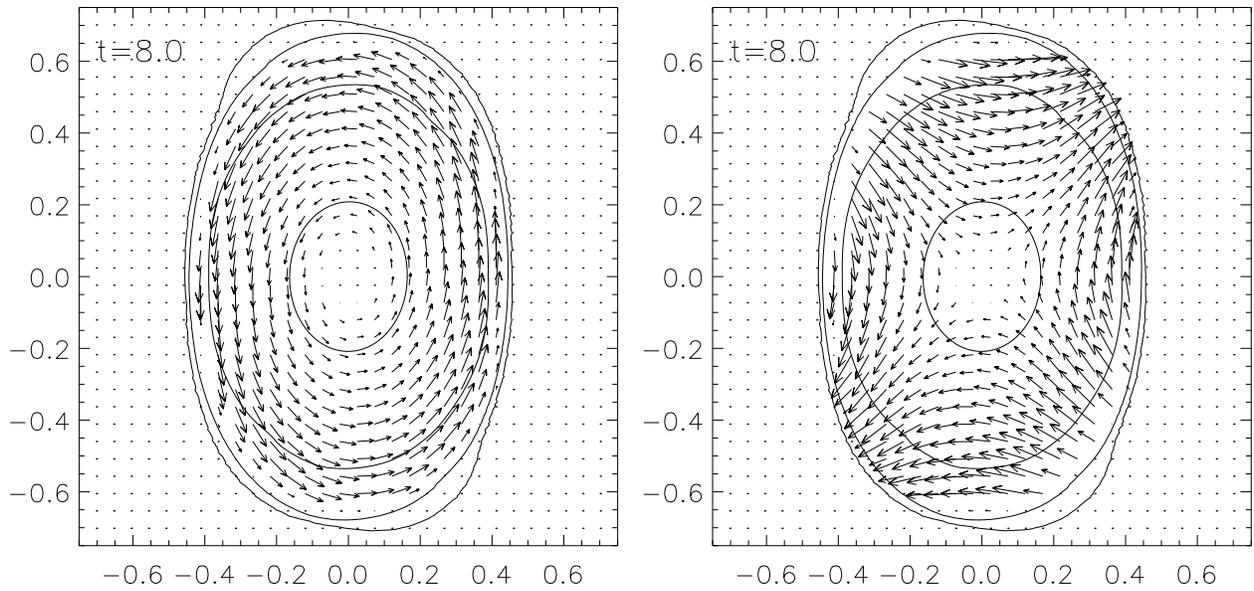} \caption{The structure of model
ROT181 is shown at time $t = 8\tau_{\mathrm{spin}}$, during the
{\it intermediate} phase of its evolution.  In both frames, solid
curves are isodensity contours in the equatorial plane while
vectors illustrate the equatorial-plane, velocity flow field as
viewed from a frame rotating with a specific frequency as follows:
$\Omega_{\mathrm{frame}}=\langle\omega_{\mathrm{r}}\rangle/2$
(left); $\Omega_{\mathrm{frame}}=\Omega_{\mathrm{rot}}$ (right).
\label{velocityVectorsEarly}}
\end{figure}

After approximately seven rotation periods, the amplitude of
$|D_{22}|$ begins to saturate, and the model deforms into a
clearly visible bar-like configuration with an axis ratio measured
in the equatorial plane of approximately 2:1 (see
Fig.~\ref{velocityVectorsEarly}). The bar-like structure is
initially spinning with a frequency given by
$\langle\omega_{\mathrm{r}}\rangle/2$, as measured during the {\it
early} phase of the ROT181 evolution.  This pattern frequency of
the bar is a factor of 7.2 smaller than the rotation frequency
$\Omega_{\mathrm{rot}}$ of the model in its initial, axisymmetric
state, so it is not surprising that the bar also exhibits sizeable
internal motions -- it has a ``Dedekind-like'' structure. Figure
\ref{velocityVectorsEarly} illustrates the structure of the model
at this time.  Both frames contain the same set of
equatorial-plane, isodensity contours delineating the bar, along
with a set of velocity vectors depicting the fluid flow inside the
bar: on the left-hand-side, the velocity vectors are drawn in a
frame corotating with the bar ({\it i.e.}, rotating at the
frequency, $\langle\omega_{\mathrm{r}}\rangle/2$) to illustrate
the elliptical streamlines of fluid flow within the
``Dedekind-like'' bar; on the right-hand-side, the velocity
vectors are drawn in a frame rotating at the frequency
$\Omega_{\mathrm{rot}}$. When viewed in this latter frame, one
sees a global velocity structure that is very similar to the
flow-field depicted in Fig.~\ref{velPerturb}, that is, it
resembles the natural eigenfunction of the $\ell = m = 2$ bar-mode
that was derived by perturbation analysis for nonrotating
spherical stars, such as our model SPH.  We note that this
velocity structure developed spontaneously in model ROT181, as the
initial model contained no velocity perturbation.

\begin{figure}
\epsscale{.90} \plotone{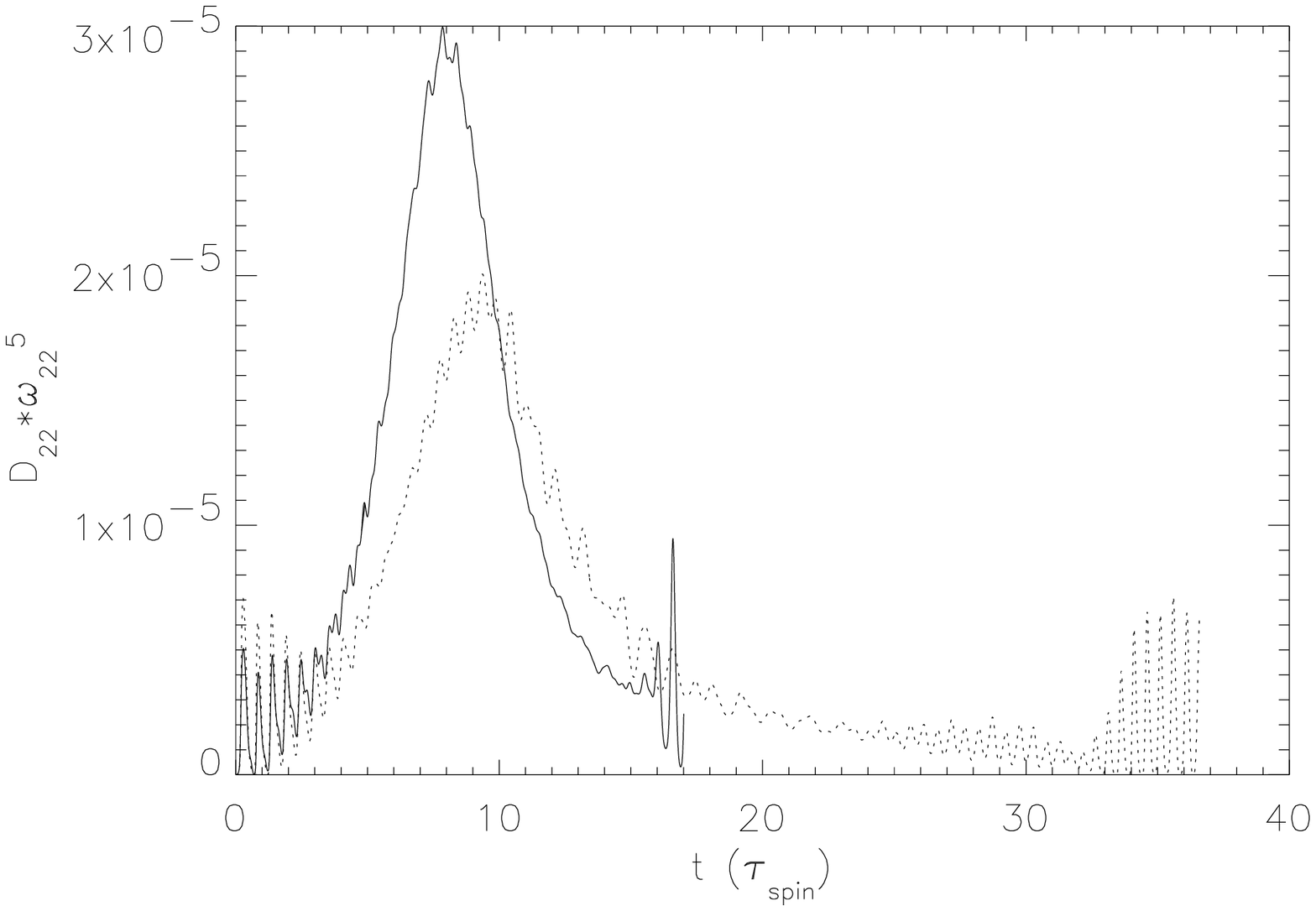} \caption{From model ROT181
(ROT179), the solid (dotted) curve depicts the time-evolution of
the product $\omega_{22}^5|D_{22}|$, which indicates the strength
of $\Phi_{\mathrm{GR}}$ in the equation of motion.  Time is shown
in units of the initial rotation period. \label{rot181d22}}
\end{figure}

During this {\it intermediate} phase of the model's evolution, the
bar remains a robust configuration, but its pattern frequency
slows as the system loses approximately 10\% of its angular
momentum (through gravitational radiation) and $T/|W|$ drops to a
value $\sim 0.156$.  It is particularly interesting to note that,
during this phase of the evolution, the GRR driving term in the
equation of motion reaches a maximum, then drops as rapidly as it
initially rose; this is illustrated in Fig.~\ref{rot181d22}, where
we have plotted the time-dependent behavior of the product,
$|\omega_{22}|^5 |D_{22}|$.  Although the bar maintains a
nonlinear structure, {\it i.e.}, $|D_{22}|$ remains large, during
this {\it intermediate} phase of the model's evolution
$\Phi_{\mathrm{GR}}$ drops quickly in concert with a decrease in
the frequency $|\omega_{22}|$.

\begin{figure}
\epsscale{1.0} \plotone{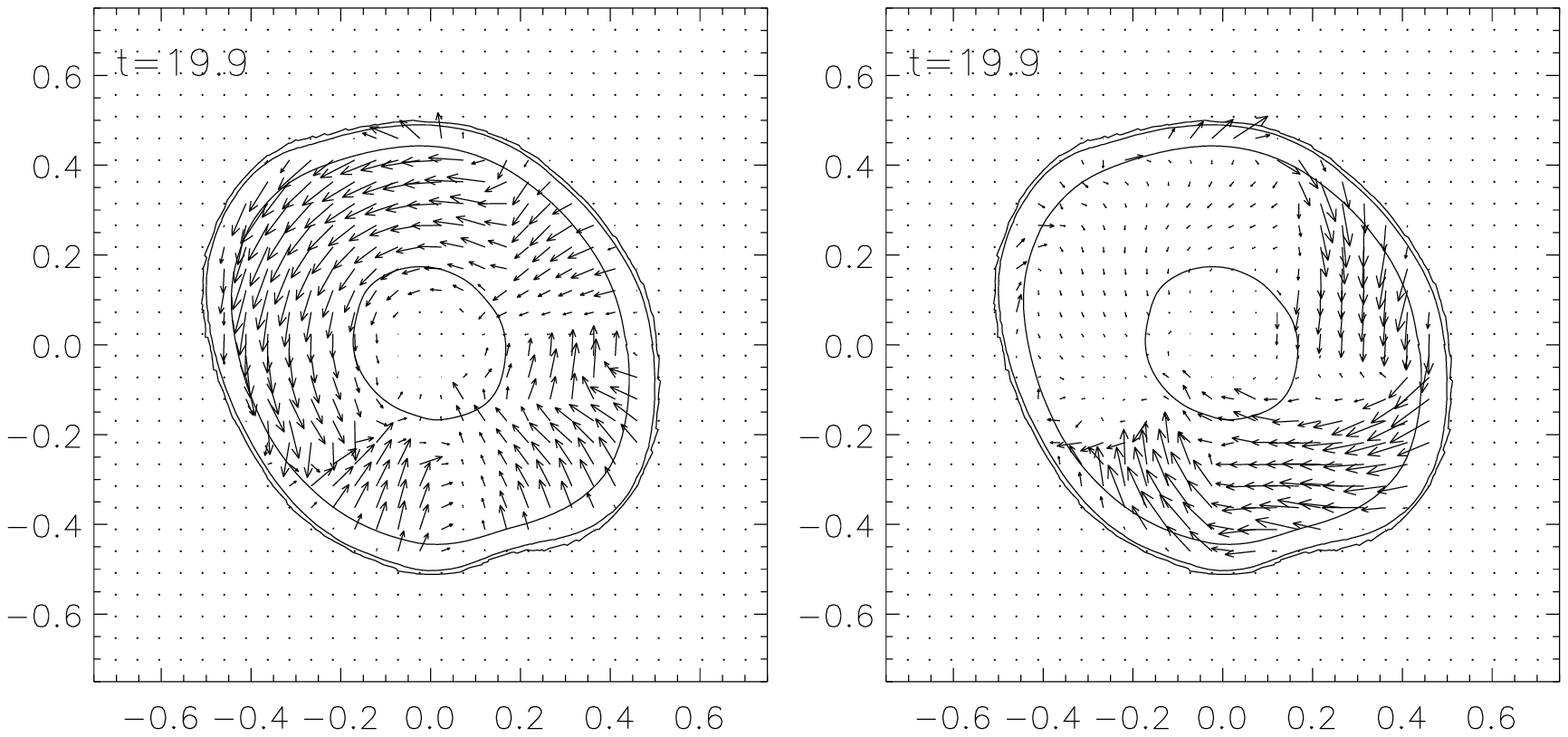} \caption{The neutron star's
structure is shown at time $t = 19.9~\tau_{\mathrm{spin}}$ during
the {\it late} phase of the ``revised'' ROT181 model evolution. In
both frames, solid curves are isodensity contours in the
equatorial plane while vectors illustrate the equatorial-plane,
velocity flow field as viewed from a frame rotating with a
specific frequency as follows: $\Omega_{\mathrm{frame}}=0$ (left);
$\Omega_{\mathrm{frame}}=\Omega_{\mathrm{rot}}$
(right).\label{velocityVectorsLate}}
\end{figure}

During the {\it late} phase of the model ROT181 evolution, the
Dedekind-like bar began to lose its coherent structure.
Small-scale fluctuations in the density and velocity fields
developed throughout the volume of the bar, and these fluctuations
grew in amplitude on a dynamical time scale.  Even vertical
oscillations developed throughout the model, disrupting both the
vertically stratified planar flow and reflection symmetry through
the equatorial plane that persisted throughout the {\it early} and
{\it intermediate} phases of the model's evolution.  After
approximately $15\tau_{\mathrm{spin}}$, the model was no longer a
recognizable bar, although it remained decidedly nonaxisymmetric,
showing density and velocity structure on a wide range of scales
in all three dimensions.  Figure \ref{velocityVectorsLate}
provides a snapshot of model ROT181's structure at
$t=19.9~\tau_{\mathrm{spin}}$ during the {\it late} phase of its
evolution. (Actually, Fig.~\ref{velocityVectorsLate} is drawn from
the {\it late} phase of a ``revised'' evolution of model ROT181,
which was evolved further in time; see \S\ref{revised} for
details.) Isodensity contours reveal a nonaxisymmetric structure
that no longer can be described simply as a bar and, when viewed
from a frame rotating at a frequency $\Omega_{\mathrm{rot}}$ (the
right-hand frame), the flow field is seen to be more complex than
in the bar.

\begin{figure}
\epsscale{.90} \plotone{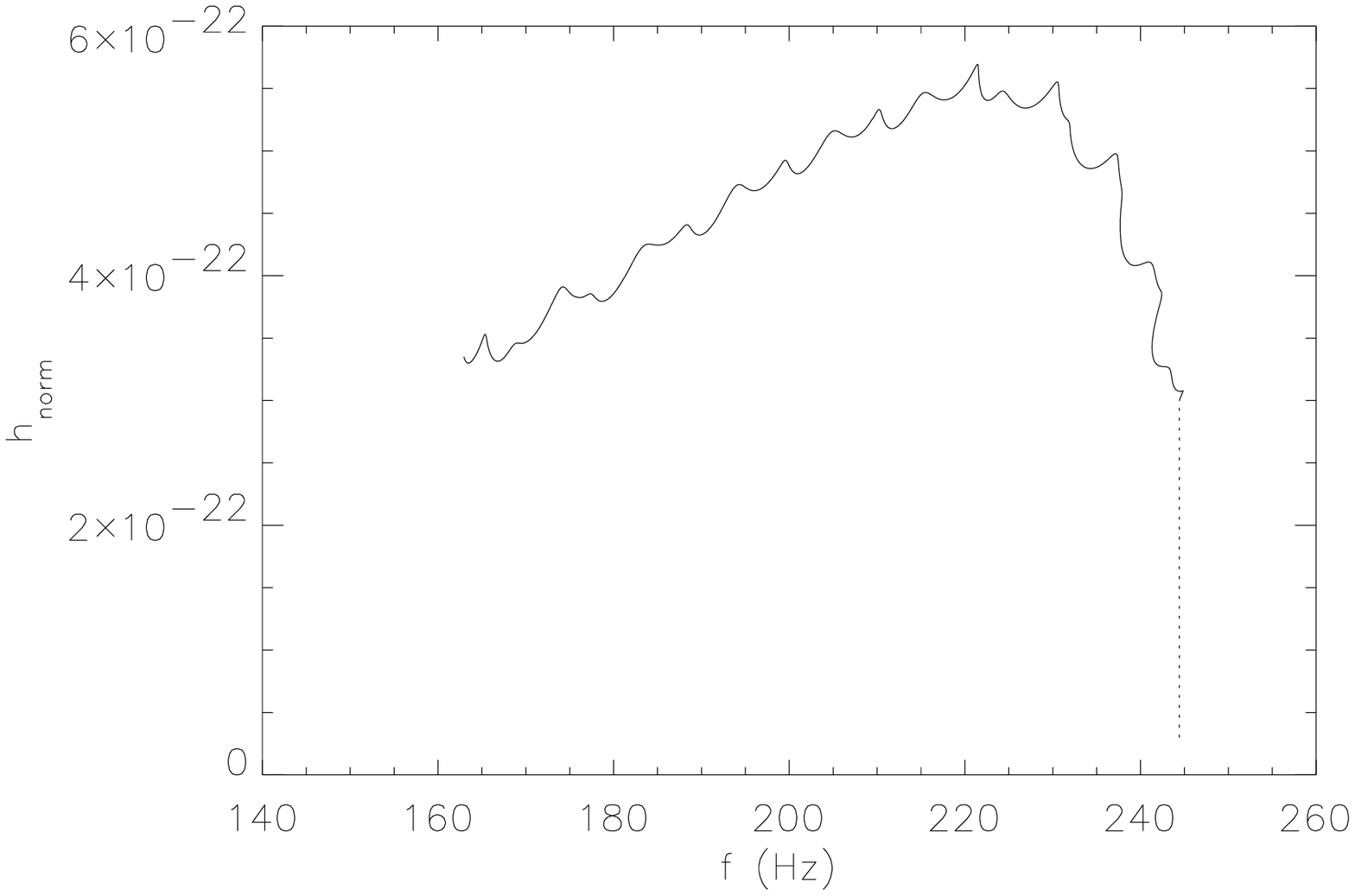} \caption{The solid curve traces
the evolution of model ROT181 in a ``strain-frequency'' diagram
from $6\tau_{\mathrm{spin}}$ to $11\tau_{\mathrm{spin}}$. As is
schematically illustrated by the vertical dotted line, initially,
the amplitude $h_{\mathrm{norm}}$ of the gravitational wave signal
grows at a constant frequency, $f
=\omega_{\mathrm{r}}/(2\pi)\approx 240~\mathrm{Hz}$. As energy and
angular momentum are radiated from the system, the frequency drops
monotonically, and the strain reaches a maximum amplitude then
steadily declines. \label{DongLaiPlot}}
\end{figure}

\subsubsection{Detectability of gravitational-wave radiation}
A rapidly spinning neutron star located in our Galaxy (and perhaps
anywhere in our local group of galaxies) that acquires the type of
nonlinear-amplitude, bar-like structure that developed in model
ROT181 will produce gravitational radiation at a frequency and
amplitude that should soon be detectable by gravitational-wave
detectors such as LIGO \citep{LIGO,S1burst}, VIRGO \citep{VIRGO},
GEO600 \citep{GEOa,GEOb}, or TAMA300 \citep{TAMA}. As our
simulation shows, however, both the amplitude and pattern
frequency of the bar --- and, hence, the strength and observed
frequency of the gravitational radiation --- will vary with time.
To illustrate this, Fig.~\ref{DongLaiPlot} depicts the evolution
of model ROT181 across a ``strain-frequency'' diagram, which is
often referenced by the experimental relativity community when
discussing detectable sources of gravitational radiation.
Specifically, the dimensionless strain
$h_{\mathrm{norm}}\equiv\sqrt{h_{+}^2+h_{\times}^2}$, where
$h_{+}$ and $h_{\times}$ are the two polarization states of
gravitational waves. For an observer located a distance $r$ along
the axis ($\theta=0$, $\phi=0$) of a spherical coordinate system
with the origin located at the center of mass of the system, we
have $h_{+} = \frac{G}{c^4} \frac{1}{r}
(\,\ddot{\!\Ibar}_{xx}-\,\ddot{\!\Ibar}_{yy})$ and $h_{\times} =
\frac{G}{c^4} \frac{2}{r} \,\ddot{\!\Ibar}_{xy}$, where the
reduced moment of intertia $\Ibar_{lm} \equiv \int \rho (x_{l}
x_{m}-\frac{1}{3}\delta_{lm}x_kx_k)dx^3$. To obtain the strain
values $h_{\mathrm{norm}}$ shown in Fig.~\ref{DongLaiPlot}, we
have assumed $r=10\mathrm{kpc}$, and the time-derivative of each
reduced moment of inertia was evaluated numerically using the
method recommended by \cite{FE90}. Model ROT181's evolutionary
trajectory in this diagram is strikingly similar to the trajectory
that was predicted by \cite{LS95} -- see their Fig.~4 -- using a
much simpler, approximate model for the development of the secular
bar-mode instability in young neutron stars.

In order to estimate the distance to which a gravitational wave
source of this type would be detectable by a gravitational-wave
interferometer, such as LIGO, we could integrate under the curve
in Fig.~\ref{DongLaiPlot}, taking into account the amount of time
that the source spends in each frequency band.   Because we have
artificially amplified the strength of the GRR force, however, our
model evolves through frequency space along the curve shown in
Fig.~\ref{DongLaiPlot} much more rapidly than would be expected
for a real neutron star that experiences this type of instability,
hence our model cannot be used directly to estimate the length of
time that such a source would spend near each frequency. However,
\cite{OL01} have outlined a method by which the detectability of a
source can be estimated from a knowledge of $\Delta J$, the total
angular momentum that is radiated away from the source via
gravitational radiation.  Specifically, the signal-to-noise ratio
$S/N$ that could be achieved by optimal filtering
can be estimated from the expression,
\begin{equation}\label{noise}
\biggl(\frac{S}{N}\biggr)^2 \approx \frac{4G}{5m\pi c^3 r^2}
\frac{|\Delta J|} {f S_h (f)} \, ,
\end{equation}
where $m$ is the azimuthal quantum number ($m=2$ for the
bar-mode), $r$ is the distance to the source, and $S_h (f)$ is the
power spectral density of the detector noise at frequency $f.$
From our model ROT181 evolution, we find $\Delta J = 1.67 \times
10^{48} \mathrm{g}~\mathrm{cm}^2 \mathrm{s}^{-1}$; and when it
reaches its design sensitivity, LIGO's $4~\mathrm{km}$
interferometer noise curve\footnote{The projected noise curve for
LIGO's $4~\mathrm{km}$ interferometers -- published as part of the
LIGO Science Requirements Document (SRD) -- and the actual noise
curve achieved by the  $4~\mathrm{km}$ interferometer at the
LIGO-Hanford Observatory (LHO) during the S3 science run can be
obtained from
\url{www.ligo.caltech.edu/$\sim$lazz/distribution/LSC\_Data/}.}
should exhibit $\sqrt{S_h}\approx 3 \times 10^{-23}
~\mathrm{Hz}^{-1/2}$ at $f=220 ~\mathrm{Hz}$, which is the
characteristic frequency of the spinning bar in model ROT181. From
expression (\ref{noise}), we therefore estimate that a source of
the type we are modelling will be detectable by LIGO with a
$S/N\gtrsim 8$ out to a distance of $2~\mathrm{Mpc}$. (During
LIGO's S3 science run in late 2003, the $4 ~\mathrm{km}$ LHO
interferometer had already come within a factor of two of this
design sensitivity.\footnotemark[1])  With Advanced LIGO (using
sapphire test masses, the projected noise curve\footnote{Projected
noise curves for the Advanced LIGO design using either sapphire or
silica test masses can be obtained from
\url{www.ligo.caltech.edu/advLIGO}.} gives $\sqrt{S_h}\approx 2.0
\times 10^{-24} ~\mathrm{Hz}^{-1/2}$ at $f=220 \mathrm{Hz}$) we
estimate that this type of source will be detectable with $S/N
\gtrsim 8$ out to $32~\mathrm{Mpc}$.

Of course the detectability of gravitational waves generated by
the secular bar-mode instability will also depend on the frequency
with which such events occur nearby.  To estimate an event rate we
can draw on the discussion of \cite{Kokkotas04} where an estimate
was made of the event rate of the dynamical bar-mode instability
in young neutron stars. Since the conditions required for the
onset of the secular bar-mode instability ($T/|W| \gtrsim 0.14$)
are almost as extreme as the conditions required for the onset of
the dynamical bar-mode instability ($T/|W| \gtrsim 0.27$), it
would be very surprising if the two event rates were not similar.
If we assume that only young neutron stars can be rotating rapidly
enough to be susceptible to either bar-mode instability, and if we
assume that a neutron star can form only from the collapse of the
core of massive star, then a reasonable upper limit on the rate of
these events will be given by the event rate of Type II
supernovae, that is, 1-2 per century per gas-rich galaxy
\citep{CET99}. (Another scenario is that rapidly rotating neutron
stars form from the accretion-induced collapse of white dwarfs.
But according to Liu 2002, the frequency of such events is orders
of magnitude lower than the event rate of Type II supernovae.)
Adopting a local galaxy density of $n_{\mathrm{g}}\approx 0.01~
\mathrm{Mpc}^{-3}$ \citep{KNST01}, we should expect $\lesssim 30$
Type II supernovae each year out to 32 Mpc. Now not all Type II
supernovae will produce neutron stars (Kokkotas 2004 estimates,
for example that 5-40\% of supernova events produce black holes
instead), and only a fraction $f_{\mathrm{rot}}$ of neutron stars
will be formed with sufficient rotational energy to be susceptible
to a bar-mode instability, so the predicted event rate should be
reduced accordingly.  A naive estimation based on angular momentum
conservation during core collapse suggests that virtually all
newly born neutron stars will be formed rapidly rotating and,
therefore, $f_{\mathrm{rot}} \sim 1$; this is the direction
\cite{Kokkotas04} leans.  But models of axisymmetric core collapse
\citep{Tohline84,DFM02a,DFM02b,OBLW04} indicate that the ratio of
energies $T/|W|$ in a newly formed neutron star is quite sensitive
to the equation of state of the core during its collapse and it is
easy to imagine physical scenarios in which appropriately rapidly
rotating neutron stars will rarely be formed; therefore,
$f_{\mathrm{rot}} \ll 1$. At the present time it is not clear
which picture is more correct, but adopting the more optimistic
view it should be possible for LIGO to detect on the order of ten
such events each year.

\subsubsection{Model Convergence}\label{revised}
\begin{figure}
\epsscale{.90} \plotone{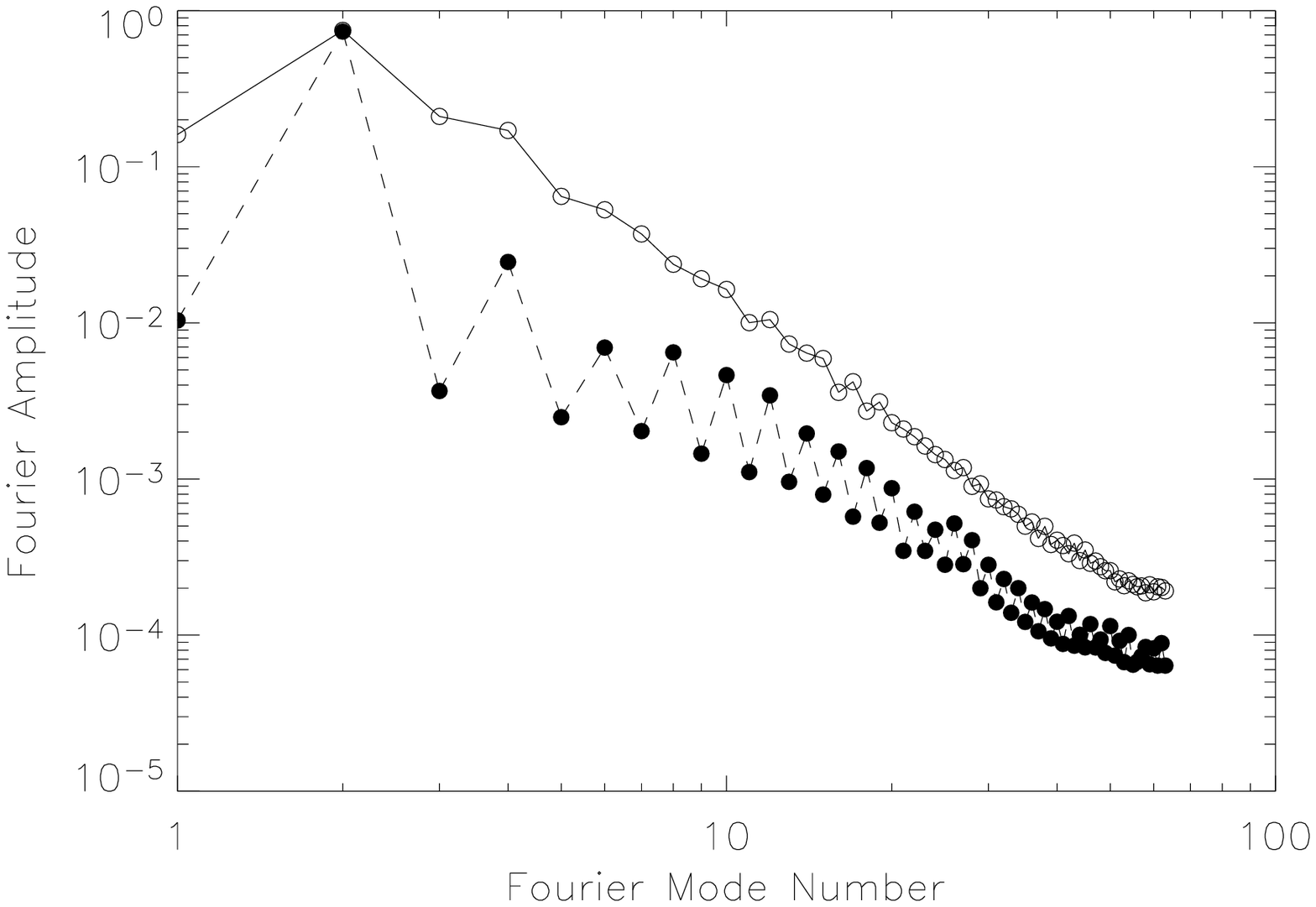} \caption{A spectrum of the
Fourier-mode amplitude of the azimuthal density distribution is
shown at time $t=10\tau_{\mathrm{spin}}$ (filled circles), when
the bar was well-developed, and at time $t=20\tau_{\mathrm{spin}}$
(open circles), after the higher-order modes destroyed the
coherent bar in the ``revised'' evolution of model ROT181. To
guide the eye, amplitudes determined for various modes at the same
time are connected by straight line segments.
\label{modeamplitudesLines}}
\end{figure}

In an effort to determine whether the Dedekind-like bar structure
was destroyed during the {\it late} phase of the ROT181 model
evolution as a result of physically realistic, hydrodynamical
processes, or by a radiation-reaction force that was artificially
too large, we set $\kappa=0$ then re-ran the last segment of the
simulation, starting from $t=11\tau_{\mathrm{spin}}$.  This
``revised'' evolution produced results that were qualitatively
identical to the late phase of the GRR-driven evolution.  That is,
the bar was destroyed by the dynamical development of velocity and
density structure on a wide range of scales in all three
dimensions.  In an effort to quantitatively describe this
relatively complex structure, Fig.~\ref{modeamplitudesLines} shows
a representation of the azimuthal Fourier-mode amplitudes of the
model's density distribution at two points in time:
$t=10\tau_{\mathrm{spin}}$, when the bar was well-developed; and
$t=20\tau_{\mathrm{spin}}$, after the higher-order nonaxisymmetric
structure was well-developed.  (Note that the {\it late} phase of
this ``revised'' evolution was followed somewhat farther in time
than the original model ROT181 evolution described in \S4.2.1.) At
the earlier time, only the $m=2$ amplitude contained a significant
amount of power, and all odd amplitudes were smaller than their
even neighbors.  At the later time the Fourier-mode amplitudes
appear to be related to one another by a simple power law,
indicating that power has been spread smoothly over all resolvable
length scales.

As an additional test of the reliability of our results, we
repeated our rotating model evolution on a computational grid that
had a factor of two poorer resolution in each of the three spatial
dimensions; specifically, the new simulation was performed on a
grid with $66\times 64\times 66$ zones in $\varpi$, $\phi$, and
$z$, respectively. On this lower resolution grid, it was not
possible to begin the evolution from precisely the same initial
state as model ROT181.  However, we were able to construct a
uniformly rotating, $n=1/2$ polytrope with $T/|W| = 0.179$ (only
$1\%$ less than the corresponding initial energy ratio of model
ROT181); we introduced a nonaxisymmetric density perturbation that
produced approximately the same initial mass-quadrupole moment
amplitude $|D_{22}|$ as in model ROT181; and throughout the
evolution the coefficient of the radiation-reaction force term in
Eq. \ref{motion} was set to $\kappa=1.75\times10^5$, as in model
ROT181. Hereafter, we will refer to this lower resolution
evolution as model ROT179.

The key results of this lower resolution evolution are illustrated
by the curves in Figs.~\ref{rot181freq} and \ref{rot181d22} that
extend beyond $\tau_{\mathrm{spin}}= 35$. As shown in the bottom
frame of Fig.~\ref{rot181freq}, the real and imaginary components
of the eigenfrequency $\omega_{22}$ identified during model
ROT179's {\it early} evolution were nearly identical to the values
measured in model ROT181; as shown in the top of
Fig.~\ref{rot181freq}, $|D_{22}|$ grew exponentially up to a
nonlinear value, levelled off, then slowly decayed as the
bar-mode's pattern frequency slowed; and as shown in
Fig.~\ref{rot181d22} the strength of the GRR force grew steadily
up to a maximum value that was somewhat lower in amplitude and
somewhat delayed in time compared to model ROT181, then almost as
rapidly it dropped by an order of magnitude, as in model ROT181.
Finally, during the {\it late} phase of model ROT179's evolution,
the bar's coherent structure was destroyed by the development of
dynamical structure on much smaller scales, just as was observed
during the {\it late} phase of model ROT181's evolution.  (This
phenomenon is evidenced in Fig.~\ref{rot181freq} by the rapid
oscillations in both components of $\omega_{22}$ at times
$t\gtrsim 32\tau_{\mathrm{spin}}$.)   In this lower resolution
evolution, however, the small-scale dynamical structure took
roughly twice as long to develop as in model ROT181. The somewhat
slower initial growth of the bar-mode and the bar's lower peak
nonlinear amplitude can both be attributed directly to this
simulation's coarser spatial resolution.  The delay in development
of the smaller scale structure was almost certainly due, in part,
to our inability to resolve structure on the smallest scales in
model ROT179, but the delay may also have occurred, in part,
because the bar, itself, was never as pronounced as in model
ROT181.  Similar behavior has been observed in simulations that
have analyzed the long-term stability of r-mode oscillations in
young neutron stars \citep{GLSSF02}.

\section{Summary and Conclusions}

Using nonrelativistic, numerical hydrodyamical techniques coupled with
a post-Newtonian treatment of GRR forces, we have simulated the
nonlinear development of the secular bar-mode instability in a rapidly
rotating neutron star.  In each simulation we have artificially
enhanced the strength of the GRR force term in the equation of motion
(by selecting values of the parameter $\kappa > 1$) in order to be
able to follow the secular development of the bar with a reasonable
amount of computing resources.  In each case however, $\kappa$ was set
to a small enough value that the amplitude of the mass-quadrupole
moment changed slowly, compared to the dynamical time scale of the
system, thus ensuring that the system as a whole remained in dynamical
equilibrium.  We first tested our simulation technique by studying the
evolution of the $\ell = m = 2$ bar-mode in a nonrotating neutron star
model (model SPH).  The developing bar-mode exhibited an azimuthal
oscillation frequency within 3\% of the frequency predicted by linear
theory, and the amplitude of the bar-mode damped (as predicted) at a
rate that was within 15\% of the rate predicted by linear theory.

Next, we evolved a rapidly rotating model (model ROT181), which
was predicted by linear theory to be unstable toward the growth of
the bar-mode. From the {\it early} ``linear-amplitude'' phase of
this model's evolution, we measured the bar-mode's azimuthal
oscillation frequency and its exponential growth rate; the values
are summarized in Table \ref{TabResults}.  The oscillation
frequency $\langle\omega_{\mathrm{r}}\rangle/\Omega_0$ was almost
an order of magnitude smaller than in model SPH, and
$\langle\omega_{\mathrm{i}}\rangle/(\Omega_0\kappa)$ was four
orders of magnitude smaller than (and had the opposite sign of)
the value measured in model SPH. Both of these frequency values
reflect the fact that model ROT181 was rotating only slightly
faster than the marginally unstable model (predicted to have
$T/|W| \approx 0.14$), in which both components of $\omega_{22}$
should be precisely zero.  We watched the unstable bar-mode grow
up to and saturate at a sufficiently large, nonlinear amplitude
that the bar-like distortion was clearly visible in two- and
three-dimensional plots of isodensity surfaces.  This nonlinear
bar-like structure persisted for several rotation periods and,
during this {\it intermediate} phase of the ROT181 model evolution, we
tracked the frequency and amplitude of the gravitational radiation
that should be emitted from the configuration due to its
time-varying mass-quadrupole moment.  Our model's evolution in a
``strain-frequency'' diagram closely matches the evolutionary
trajectory predicted by \cite{LS95}, lending additional
credibility to their relatively simple (and inexpensive) way of
predicting the evolution of such systems as well as to our first
attempt to model such an evolution using nonlinear hydrodynamical
techniques. During the {\it late} phase of our model ROT181
evolution, the bar lost its coherent structure and the system
evolved to a much more complex nonaxisymmetric configuration. The
general features of this {\it late} phase of the evolution were
reproduced when the simulation was re-run on a coarser
computational grid, and even when the GRR forces were turned off.
So while the size and shape of the {\it intermediate} phase
``Dedekind-like'' structure of our model may well have been
influenced strongly by the excessive strength of the GRR force
used in our simulation, it appears as though the final complex
``turbulent'' phase of the evolution was governed by purely
hydrodynamical phenomena.

It is not clear what physical mechanism was responsible for the
development of the small-scale structure and subsequent
destruction of the bar during the {\it late} phase of the
evolution of model ROT181. Because the bar's structure was
``Dedekind-like'' -- that is, fluid inside the bar was moving
along elliptical streamlines with a mean frequency that was
significantly higher than the bar pattern frequency -- it is
tempting to suggest that the small-scale structure arose due to
differential shear.  But, according to \cite{HBW99}, coriolis
forces are able to stabilize differentially rotating,
astrophysical flows against shearing instabilities even in
accretion disks where the shear is much stronger than in our
``Dedekind-like'' bar.  (See, however, Longaretti 2002 for an
opposing argument.)  Furthermore, other models of differentially
rotating astrophysical bars \citep{CT00,NCT00} do not appear to be
susceptible to the dynamical instability that destroyed the bar in
our ROT181 model evolution.  We suspect, instead, that the
late-time behavior of model ROT181 results either from nonlinear
coupling of various oscillatory modes within the star, or from an
``elliptic flow'' instability similar to the one identified in
laboratory fluids that are forced to flow along elliptical
streamlines. The dissipative effect of mode-mode (actually,
three-mode) coupling has been examined in depth by \cite{SAFTW02}
and \cite{AFMSTW03} in the context of the r-mode instability in
young neutron stars, and \cite{BTW04} have shown the connection
between this process and the rapid decay of the r-mode in extended
numerical evolutions such as the ones performed by \cite{GLSSF02}.
However, this phenomenon has not yet been studied to the same
degree in relation to the $\ell=m=2$ f-mode. \cite{LL93},
\cite{LL96}, and \cite{LS99} have demonstrated that the ``elliptic
flow'' instability seen in laboratory fluids is likely to arise in
self-gravitating ellipsoidal figures of equilibrium, especially if
they have ``Dedekind-like'' internal flows.  Additional analysis
and, very likely, additional nonlinear simulations will be
required before we are able to determine which (if either) of
these mechanisms was responsible for the destruction of the bar in
our ROT181 model evolution.

Our nonlinear simulation of model ROT181 demonstrates that when a
rapidly rotating neutron star becomes unstable to the secular
bar-mode instability, the bar-like distortion may grow to
nonlinear amplitude and thereby become a strong source of
gravitational radiation. However, it will not be a long-lived
continuous-wave source, as one might optimistically have expected;
in our simulation, the nonlinear-amplitude bar survived fewer than
ten rotation periods.  In a real neutron star the GRR forces will
be much weaker than those of our simulation, so we expect the bar
mode to grow and persist for many more rotation periods.  However,
we also expect the amplitude of the bar mode to saturate at a much
lower amplitude in a real neutron star.  Nevertheless, we expect
the bar mode to persist in rapidly rotating neutron stars long
enough to allow gravitational radiation to remove sufficient
angular momentum for them to relax into a secularly stable
equilibrium state.  Thus the amount of angular momentum radiated
away in real neutron stars should be comparable to that in our
simulation.  While such astrophysical systems may not be the
easiest sources to detect with broadband, gravitational-wave
detectors such as LIGO because the frequency of the emitted
radiation will change steadily with time, our estimates suggest
that gravitational waves arising from the excited secular bar-mode
instability in rapidly rotating neutron stars could well be
detectable in the not too distant future from neutron stars as far
away as $32~\mathrm{Mpc}$.

After submitting this paper
for publication, we became aware that \cite{SK04} have just
completed an investigation similar to the one presented here in
which they have utilized post-Newtonian simulations to study the
nonlinear development of the secular bar-mode instability in
rapidly rotating neutron stars.  Their initial models were
differentially rotating, $n=1$ ($\Gamma=2$) polytropes with $0.2
\lesssim T/|W| \lesssim 0.26$.  The {\it early} and {\it
intermediate} phases of their model evolutions agree well with the
results of our model ROT181 evolution, that is, the bar-mode grew
exponentially at rates consistent with the predictions of linear
theory and reached a nonlinear amplitude, producing an ellipsoidal
star of moderately large ellipticity.  
The strength of the GRR force used in our simulations was considerably
larger than theirs.  This may explain why the bar mode grows to a
larger amplitude and why, in turn, there is a more significant decrease
in the pattern frequency of the bar as it evolves toward a Dedekind-like
configuration in our simulation.  This may also explain why the bar
mode structure was ultimately destroyed by short wavelength disturbances
in our evolutions while such turbulence had not yet developed in theirs.

%However, they did not
%follow their evolutions far enough in time to observe a
%significant decrease in the pattern frequency of the bar toward a
%Dedekind-like configuration or to examine the late-time stability
%of the bar against short wavelength disturbances.


%% Included in this acknowledgments section are examples of the
%% AASTeX hypertext markup commands. Use \url without the optional [HREF]
%% argument when you want to print the url directly in the text. Otherwise,
%% use either \url or \anchor, with the HREF as the first argument and the
%% text to be printed in the second.

\acknowledgments

We thank Gabriela Gonz\'alez, Peter Saulson, Peter Fritschel and
Kip Thorne for guidance in obtaining the LIGO noise figures used
in our analysis. We also thank an anonymous referee for recommending
several ways in which the presentation of our results could be improved.
This work was supported in part by NSF grants
AST-9987344, AST-0407070, PHY-0326311 and NASA grant NAG5-13430 at
LSU; and by NSF grants PHY-0099568, PHY-0244906 and NASA grants
NAG5-10707, NAG5-12834 at Caltech. Most of the simulations were
carried out on SuperMike and SuperHelix at LSU, which are
facilities operated by the Center for Computation and Technology
whose funding largely comes through appropriations by the
Louisiana state legislature.




\clearpage

%% The following command ends your manuscript. LaTeX will ignore any text

\end{document}